\documentclass[aps,prl,twocolumn,showpacs,groupedaddress]{revtex4} 
\usepackage{graphicx} 
 
\begin{document} 
 
\title{Resolving the wave-vector in negative refractive media: The sign of $\sqrt{Z}$} 
\author{S. Anantha Ramakrishna} 
\affiliation{Department of Physics, Indian Institute of Technology, Kanpur - 208 016, India}
\author{Olivier J.F. Martin}
\affiliation{Nanophotonics and Metrology Laboratory, Swiss Federal Institute of Technology Lausanne, EPFL-STI-NAM, 1015 Lausanne, Switzerland}
\date{\today} 
 
\begin{abstract} 
We address the general issue of resolving the wave-vector in complex electromagnetic media including negative refractive media. This requires us to make a physical choice for the sign of a square-root imposed merely by conditions of causality. By considering the analytic behaviour of the wave-vector in the complex plane, it is shown that there are a total of eight physically distinct cases in the four quadrants of two Riemann sheets.

\end{abstract} 
\maketitle 
 
The electromagnetic response of a homogeneous medium is usually charecterised by
the dielectric constant ($\varepsilon$) and the magnetic permeability ($\mu$).
The refractive index ($n$) first arises in the context of the wave equation
and is defined by $n^2 = \varepsilon \mu$. For most optical media, $\mu=1$ and
$n$ is taken to be $\sqrt{\varepsilon}$.
These are in general complex functions (i.e., $\varepsilon = \varepsilon' +
i \varepsilon''$ and $\mu = \mu' + i\mu''$) of the frequency ($\omega$) 
of the applied 
radiation. For media at thermodynamic equilibrium, it is usually demanded 
that the imaginary parts of $\varepsilon(\omega)$ and $\mu(\omega)$ are 
positive so that the total energy dissipated by the electromagnetic fields 
 in a volume ($V$),
\begin{equation}
\int_V d^3 r \int_{-\infty}^{\infty} \omega
[\varepsilon''(\omega)\vert \vec{E}(\vec{r},\omega) \vert^2 + 
\mu''(\omega) \vert \vec{H}(\vec{r},\omega) \vert^2 ] \frac{d\omega}{2\pi}~,
\end{equation}
is positive\cite{landau}. However, the signs of the real parts of 
$\varepsilon$ and $\mu$ are not subject to any such restriction\cite{landau}.
Veselago\cite{veselago} had concluded that a material with real and 
simultaneously negative $\varepsilon$ and $\mu$ at a given frequency would 
have a negative refractive index $n = -\sqrt{\varepsilon \mu}$. This result 
remained an academic curiosity until recently when it became possible to 
fabricate structured meta-materials with negative $\varepsilon$ and $\mu$ 
\cite{smith00,smith01,eleftheriades,parazzoli,houck}.
The possibility that negative refractive index may open a door to 
perfect lenses\cite{pendry_PRL00}, not subject to the 
diffraction limit has given a great impetus to the study of these materials. 

The sign of the refractive index in these media has been subject of some 
debate. Smith and Kroll\cite{smith_kroll} analysed the problem of a 
current sheet radiating into a medium with negative $\varepsilon$ and negative
$\mu$, and concluded that $n < 0$ for power to flow away from the source. 
This has been criticised by some authors\cite{valanju,pokrovsky}.
However, there exists a very real problem of choosing the sign of the 
square-root to determine the wave-vector of a wave transmitted into a 
negative medium.  For the 
case of normally incident propagating waves with no tranverse components,
this sign of the wave-vector directly corresponds to the sign of the refractive index.

Here we consider this problem of choosing the correct wave-vector in media with 
complex $\varepsilon$ and $\mu$. Using the theory of analytic functions we show
that there is, indeed, a physical choice of the sign to be made. We include the
cases of propagating and evanescent waves, and  that where both the real and the imaginary
parts of $\varepsilon$ and $\mu$ could be positive or negative. Negative
imaginary parts are possible in an amplifying medium (as in a laser). Even for
the case of passive meta-materials which are overall only dissipative, 
it appears possible to have a negative imaginary
part of $\varepsilon$ when the real part of $\mu$ is negative\cite{obrien} 
and a negative imaginary part of $\mu$ when the real part of $\varepsilon$ is
negative\cite{koschny}. 
We show that there are a total of eight distinct physical cases in two 
Riemann sheets for the square root operation.  Our results support the case
that when a medium has predominantly real and simultaneously negative 
$\varepsilon$ and $\mu$ at a given frequency, it must be considered to have a 
negative refractive index. 

Let us examine the choice of the wave-vector in homogeneous 
media. The propagation of light in a medium is governed by Maxwell's equations. For a time-harmonic plane wave, $\exp[i(\vec{k}\cdot\vec{r}-\omega t)]$, with an angular frequency $\omega$ and a wave-vector $\vec{k}$, these reduce to 
\begin{equation}
\vec{k}\times\vec{E} = \frac{\omega}{c} \mu \vec{H}, ~~~~~~~~ 
\vec{k}\times\vec{H} = -\frac{\omega}{c} \varepsilon \vec{E},
\end{equation}
where $\vec{E}$ and $\vec{H}$ are the electric and magnetic fields 
associated with the wave. 
For complex $\varepsilon$, $\mu$ and $\vec{k}$, the wave becomes inhomogeneous. 
Writing $\varepsilon = \varepsilon '+i \varepsilon ''$, and $\mu = \mu '+i \mu''$, 
we note that the medium is absorbing if $\varepsilon ''>0 $, $\mu'' > 0$, and 
amplifying if $\varepsilon ''<0$, $\mu'' < 0$.
The mechanism for the absorption or amplification, of course, lies in the 
underlying electric or magnetic polarizabilities. We should mention here 
that a meta-material can exhibit resonances unrelated to the underlying 
material polarization, for example, an LC resonance for the split-ring 
resonators\cite{smith00,obrien}. But these can also be subsumed into an effective macroscopic 
$\varepsilon$ and $\mu$ when the wavelength of the radiation is much larger 
than the underlying structure. Then the structure appears homogeneous.
In an isotropic medium, Maxwell's equations require 
\begin{equation}
|\vec{k}|^2 = k_x^2 + k_y^2 + k_z^2 = \varepsilon\mu \frac{\omega^2}{c^2},
\end{equation}
which describes the dispersion in the medium. To determine the 
wave-vector $\vec{k}$ in the medium, we have to carry out a square root 
operation. Obviously the choice of the sign of the square root will have to 
be made so as to be consistent with the Maxwell's equations and causality.  

\begin{figure}[tbp] 
\includegraphics[width=8cm]{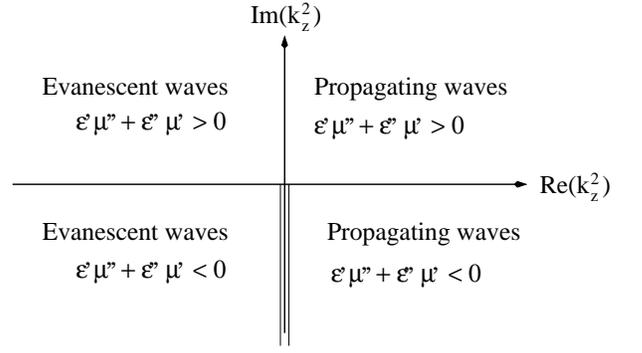} 
\caption{The complex plane for $k_z^2$ showing the different regions for the propagating and evanescent waves. The branch cut along the negative imaginary axis for the square root is shown.}
\label{complexplane1} 
\end{figure}

Without loss of generality, we consider an electromagnetic wave with a 
wave-vector $[k_x,~0,~k_z]$ to be incident from vacuum on the left
($-\infty<z<0$) on a semi-infinite medium ($\infty > z >0$) with an arbitrary 
value of $\varepsilon$ and $\mu$. Due to $x$-invariance, $k_x$ is preserved 
across the interface. The $z$-component of the wave-vector,$k_z$, however, 
has to be obtained from the dispersion relation:
\begin{equation}
k_z = \pm \sqrt{\varepsilon\mu \frac{\omega^2}{c^2} - k_x^2},
\label{w_vec}
\end{equation}
where a physical choice has to be made for the sign of the square root.  
Now the waves in medium-2 could be propagating [$ k_x^2 < \mathrm{Re}
(\varepsilon\mu \omega^2/c^2)$] or evanescent [$k_x^2 > \mathrm{Re}(\varepsilon
\mu \omega^2/c^2)$]. Further the media could be absorbing or amplifying 
depending on the sign of $\mathrm{Im}(\varepsilon \mu)$ in Eqn.~(\ref{w_vec}). 
This enables us to divide the complex plane for $Z = k_z^2$ into the four 
quadrants shown in Fig. \ref{complexplane1}.
The waves corresponding to quadrants-1 and 4 have a propagating nature, and 
the waves corresponding to quadrants-2 and 3 are evanescent. Crucially, we 
note that there is a branch cut in the complex plane for $\sqrt{Z} = k_z$ 
and one cannot analytically continue the behaviour of the waves across this 
branch cut. This branch cut divides the Riemann surface into two sheets in which
the two different signs for the square root will have to be taken\cite{complex_analysis}.
We now have to place this branch-cut in the complex plane so as 
to be consistent  with physical boundary conditions. For absorbing media, 
the wave amplitude at the infinities has to obviously disappear. For 
amplifying media, one has to be more careful. The only conditions are that 
evanescently decaying waves remain decaying, propagating ones remain 
propagating and no information can flow in from the infinities. This ensures 
that the near-field features of a source cannot be probed at large distances 
merely by imbedding the source in an amplifying medium. Due to the above 
reasons 
we will show that the branch cut has to be chosen along the negative 
imaginary axis as shown in Fig. 1 below. Hence our range 
for the argument $\theta$ of $k_z^2$ becomes $-\pi /2 < 
\theta < 3\pi /2$  for the first Riemann sheet and $3\pi /2 < \theta < 7\pi /2$ for the second Riemann sheet, corresponding to the two signs of the square 
root $k_z = \pm \sqrt{Z} = |Z|^{1/2} e^{i\theta/2}$ and $|Z|^{1/2} e^{i\pi 
+i\theta/2}$ respectively. 
The reason for our choice becomes clear from the 
discussion below of the behaviour of the waves in different regions of 
the complex plane.

\begin{figure}[tbp] 
\includegraphics[width=8cm]{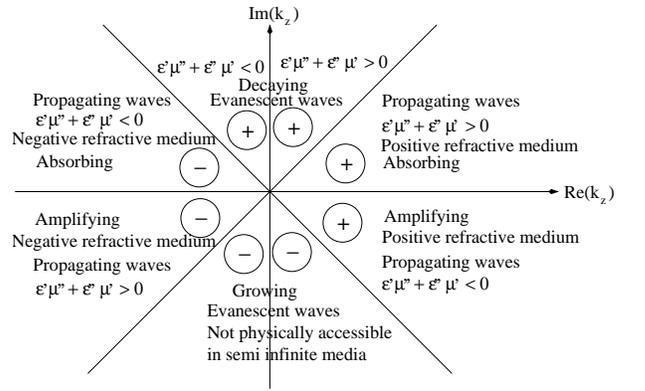} 
\caption{The complex plane for $k_z = \sqrt{Z}$ with the corresponding eight regions in the two Riemann sheets for $k_z^2$. The evanescent regions could be absorbing or amplifying as explained in the text depending on the sign of $\varepsilon'$ and $\mu'$. The sign of the square root for the region is shown enclosed in a small circle.}
\label{complexplane2} 
\end{figure}
The complex plane for $k_z = \sqrt{Z}$ with the corresponding eight regions in the two Riemann sheets is shown in Fig. 2. We will now consider the behaviour of the waves individually in each of these regions.\\ 
\noindent{\it Region-1}: $0<\mathrm{Arg}(k_z) < \pi/4$ corresponding to $0 < \mathrm{Arg}(k_z^2) < \pi/2$ --
This is the conventional case of a propagating wave in a positively refracting absorbing medium. Here $\mu'\varepsilon'' + \varepsilon' \mu'' > 0$ immplying
absorption.  The wave decays in amplitude as it propagates in  the medium. We also have $\varepsilon' > 0$ and $\mu'>0$ and choose the positive sign of the square root.\\
{\it Region-2}: $\pi/4<\mathrm{Arg}(k_z) < \pi/2$ corresponding to $\pi/2 < \mathrm{Arg}(k_z^2) < \pi $ --
This is a case of evanescently decaying waves. We have $\varepsilon' \mu''+\varepsilon'' \mu' > 0 $ which implies overall absorption, for example, 
 ($\varepsilon''>0$, $\mu''>0$)  if $\varepsilon'> 0$, $\mu'>0$ 
(positive refractive index) or amplification ($\varepsilon''<0$, $\mu''<0$) 
if $\varepsilon'< 0$, $\mu'<0$ (negative refractive index). Note that it is 
the overall sign of $\varepsilon' \mu''+\varepsilon'' \mu'$ that matters. In either case we have an evanescent wave that decays into the medium to zero as $z \rightarrow \infty$.\\
{\it Region-3}: $\pi/2<\mathrm{Arg}(k_z) < 3\pi/4$ corresponding to $\pi < \mathrm{Arg}(k_z^2) < 3\pi /2 $ --
This is similiar to region-2 and corresponds to evanescently decaying waves. Although we have now that $\varepsilon' \mu''+\varepsilon'' \mu' < 0 $, implying absorption for negative refractive media and amplification for positive refractive media. Again we choose the positive square root and we only have evanescently decaying waves in the semi-infinite medium. \\
{\it Region-4}: $3\pi/4<\mathrm{Arg}(k_z) < \pi$ corresponding to $3\pi /2 < \mathrm{Arg}(k_z^2) < 2 \pi $ --
Now we have moved into the second Riemann sheet and choose the negative sign for the square root. This region corresponds to negatively refracting media ($\varepsilon'< 0$, $\mu'<0$) and absorbing media ($\varepsilon' \mu''+\varepsilon'' \mu' < 0 $) and hence, for example, $\varepsilon''>0$, $\mu''>0$. We have propagating waves (albeit left-handed) that decay in amplitude as the wave propagates into the medium. Note that the negative sign is crucial to ensure this decaying nature of the waves in absorbing media.\\
{\it Region-5}: $\pi<\mathrm{Arg}(k_z) < 5\pi /4$ corresponding to $2\pi < \mathrm{Arg}(k_z^2) < 5 \pi/2 $ --
We again have the negative sign for the square root and propagating waves in negative index media. But note that $\varepsilon' \mu''+\varepsilon'' \mu' > 0 $ implying, for example, that $\varepsilon''<0$, $\mu''<0$. In general, overall there is amplification and the waves grow exponentially with propagation distance.\\
{\it Region-6}: $5\pi /4<\mathrm{Arg}(k_z) < 3\pi /2$ corresponding to $5\pi/2 < \mathrm{Arg}(k_z^2) < 3 \pi$ --
This corresponds to exponentially growing evanescent waves which are not physically accessible in semi-infinite media. For finite media (slabs) these solutions are permissible and are responsible for the perfect lens effect\cite{pendry_PRL00}.\\
{\it Region-7}: $3\pi /2<\mathrm{Arg}(k_z) < 7\pi /4$ corresponding to $3\pi  < \mathrm{Arg}(k_z^2) < 7 \pi/2$ --
This also corresponds to exponentially growing evanescent waves and are not physically accessible in semi-infinite media.  \\
{\it Region-8}: $-\pi /4<\mathrm{Arg}(k_z) < 0$ corresponding to $-\pi /2  < \mathrm{Arg}(k_z^2) < 0$ --
Now we are back on the first Riemann sheet and choose the positive sign for the square root. This is the conventional case of propagating waves in amplifying postive refractive media which grow exponentially in amplitude into the medium. 

Note that in all these cases it is not $\varepsilon$ or $\mu$ that individually determine the nature of the waves, but a combination of them determined by $\mathrm{Re} (k_z^2)$ and $\mathrm{Im}( k_z^2)$. For evanescent waves in a semi-infinite medium, we always choose the positive square root so that $\mathrm{Im} (k_z) \ge 0$. In the case of evanescent waves in amplifying media, our choice results in a Poynting vector in the medium that points towards the source (interface in this case). This, however, does not violate causality as the Poynting vector/energy flow decays exponentially to zero at infinity and no information flows in from the infinities. This counter-intuitive behaviour does not imply that source has turned into a sink - rather it indicates that there would be a large (infinitely large for unsaturated linear gain) accumulation of energy density (intense local field enhancements) near a source.  This behaviour can also  be understood in terms of the fundamental Bosonic property of light - photons in the localized  mode stimulate the amplifying medium to emit more photons into the same localized  mode. In other words, the near-field modes of a source remain evanescent even inside an amplifying medium and do not affect the far-field.
Also note that $\varepsilon'$ and $\mu'$ (such as in a metal) could have opposite signs in which case the waves are evanescent and fall into regions-2 or 3 depending on the overall sign of $\varepsilon'\mu''+\varepsilon''\mu'$. 
 
In summary, the problem of choosing wave-vector in complex media can be 
analysed using the properties of analytic functions. The branch-cut  
for the square root in the complex plane of 
$Z = k_z^2$  has to be chosen along the negative imaginary axis. This results in
eight distinct cases for the quadrants in two Riemann sheets corresponding 
variously to propagating or evanescent waves in absorbing or amplifying media, 
with positive or negative real parts of $\varepsilon$ and $\mu$. 
The positive sign
of the square root has to be taken for the cases in one sheet and the negative
sign in the other. The case of propagating waves in absorbing media with 
$\varepsilon' <0$ and $\mu' <0$ lies in the second sheet which justifies 
calling them negative refractive index media.

\end{document}